\documentclass[twocolumn]{aa}  
\usepackage{graphicx}
\usepackage{color}
\usepackage{epsfig}
\usepackage{supertabular}
\usepackage{aalongtable}
\usepackage{txfonts}
\begin{document}
   \title{A search for ultra-compact dwarf galaxies in the Centaurus galaxy cluster\thanks{Based on observations obtained in service mode at the VLT (programme 076.B-0293)}}


   \author{Steffen Mieske
          \inst{1}
          \and
          Michael Hilker\inst{1}\and Andr\'{e}s Jord\'{a}n\inst{1} \and Leopoldo Infante\inst{2} \and Markus Kissler-Patig\inst{1}
          }

   \offprints{S. Mieske}

   \institute{European Southern Observatory, Karl-Schwarzschild-Strasse 2, 85748 Garching bei M\"unchen, Germany\\
              \email{smieske@eso.org;mhilker@eso.org;ajordan@eso.org;mkissler@eso.org}
         \and
     Departamento de Astronom\'{\i}a y Astrof\'{\i}sica, Pontificia
Universidad Cat\'olica de Chile, Casilla 306, Santiago 22, Chile\\
\email{linfante@astro.puc.cl}
}

   \date{}

 
  \abstract
   {} 
   {Our aim is to extend the investigations of ultra-compact dwarf galaxies (UCD) beyond the well  studied Fornax and Virgo clusters.}
   {We measured spectroscopic redshifts of about 400 compact object
     candidates with $19.2<V<22.4$ mag in the central region of the
     Centaurus galaxy cluster (d=43Mpc), using 3 pointings with
     VIMOS@VLT.  The luminosity range of the candidates covers the
     bright end of the globular cluster (GC) luminosity function and
     the luminosity regime of UCDs in Fornax and Virgo. Within the area
     surveyed, our completeness is $\approx$30\%.}
   {We find 27 compact objects with radial velocities consistent with
     them being members of Centaurus, covering an absolute magnitude
     range $-12.2<M_V<-10.9$ mag. We do not find counterparts to the
     two very large and bright UCDs in Fornax and Virgo with
     $M_V=-13.5$ mag, possibly due to survey incompleteness. The
     compact objects' distribution in magnitude and space is
     consistent with that of the GC population.  Their kinematics and
     spatial distribution indicate that they are associated more to
     the central galaxies than to the overall cluster potential. The
     compact objects have a mean metallicity consistent with that of
     the metal-rich globular cluster sub-population. Compact objects
     with high S/N spectra exhibit solar [$\alpha$/Fe] abundances,
     consistent with typical dwarf elliptical galaxy values and unlike
     galactic bulge globular clusters.  HST based size estimates for a
     sub-sample of eight compact objects reveal the existence of one
     very large object with half-light radius $r_h$ around 30 pc,
     having $M_V=-11.6$ mag ($\simeq 10^7$ M$_{\sun}$).  This source
     shows super-solar [$\alpha$/Fe] abundances.  Seven further
     sources are only marginally larger than typical GCs with $r_h$ in
     the range 4 to 10 pc. Those sources exhibit a large scatter in
     [$\alpha$/Fe] abundances.}
   {We consider the largest compact object found to be the only
     bona-fide UCD detected in our study. In order to improve our
     understanding of UCDs in Centaurus, a significant increase of our
     survey completeness is necessary. }


   \maketitle 
%

\section{Introduction}
\label{Intro}
In their spectroscopic studies of the Fornax galaxy cluster, Hilker et
al.  (\cite{Hilker99}) and Drinkwater et al. (\cite{Drinkw00}
and \cite{Drinkw03}) reported on the discovery of six isolated compact
stellar systems, having $-13.5<M_V<-12$ mag and half-light radii $r_h$
between 20 and 100 pc. Due to their compactness compared to dwarf
galaxies of similar luminosity, they were dubbed ``ultra-compact dwarf
galaxies`` (UCDs) (Phillipps et al.~\cite{Philli01}).  More recently, UCDs
have also been discovered in the Virgo cluster (Ha\c{s}egan et
al.~\cite{Hasega05}, Jones et al.~\cite{Jones05}).  

There are,
currently, two main hypotheses on the origin of UCDs: 1. UCDs are
remnant nuclei of dwarf elliptical galaxies stripped in the potential
well of their host cluster (e.g. Bassino et al.~\cite{Bassin94}, Bekki
et al.~\cite{Bekki03}).  2. UCDs are merged stellar super-clusters
created in gas-rich galaxy mergers (e.g.  Fellhauer \&
Kroupa~\cite{Fellha02}).

High resolution imaging and spectroscopy (Drinkwater et
al.~\cite{Drinkw03}, Hilker et al.~\cite{Hilker07}) place the Fornax
UCDs between the sequence of globular clusters (GCs) and dwarf
elliptical galaxies (dEs) in the fundamental plane of stellar systems.
Their M/L ratios are in the range 2-5 and can be explained without
invoking the presence of dark matter. The luminosities, sizes and M/L
ratios of Fornax UCDs are comparable to those of bright nuclei of
nucleated dwarf ellipticals (dE,Ns) (Lotz et al.~\cite{Lotz04},
C\^{o}t\'{e} et al.~\cite{Cote06}, Hilker et al.~\cite{Hilker07}). 

For the Virgo cluster, Ha\c{s}egan et al.~(\cite{Hasega05}) report on
three compact objects with M/L ratios between 6 and 9, suggesting such
high ratios as a criterion to separate UCDs from GCs. From their HST
based size estimates they furthermore derived a limit between normal
GCs and larger sources (Dwarf Globular Transition Objects in their
notation) to occur at about 2$\times 10^6$ M$_{\sun}$.  Evstigneeva et
al.~(\cite{Evstig07}) find that Virgo UCDs are in general more
$\alpha$ enriched than dwarf elliptical galaxies in Fornax and Virgo.
This is not naturally explained by the hypothesis that Virgo UCDs are
dominated by stripped nuclei.

Our long term aim is to characterise UCD properties in a broad range of
environments (e.g. Mieske et al.~\cite{Mieske04b}, Mieske et
al.~\cite{Mieske06c}). For the Fornax cluster (Mieske et
al.~\cite{Mieske02},~\cite{Mieske04a}) we have shown that the
luminosity distribution of compact objects for $M_V<-10.4$ mag does
not show any discontinuities that would hint at two separate
populations (i.e. UCDs vs. GCs)\footnote{we refer as ``compact
  objects'' to sources in the magnitude regime of bright globular
  clusters and Fornax UCDs. Only if the properties of a compact object
  cannot be explained by the general globular cluster population, we
  refer to it as UCD. In our usage the term UCD does not imply a
  particular formation channel.}. Only the brightest UCD with
$M_V=-13.5$ mag clearly stands out.  In Mieske et
al.~(\cite{Mieske06a}), we found a break in the metallicity
distribution at $M_V=-11$ mag, such that brighter compact objects are
more metal rich than fainter compact objects. Furthermore, we
identified a change in the size-luminosity relation occurring at about
the same luminosity, similar to the Virgo cluster case. Fainter
sources have luminosity independent half light radii of $r_h\simeq
3$pc (e.g. Jord\'an et al.~\cite{Jordan05}), brighter sources have
sizes correlating with their luminosity (Ha\c{s}egan et
al.~\cite{Hasega05}, Kissler-Patig et al.~\cite{Kissle05}).  We
therefore identified $M_V=-11$ mag ($\sim$3$\times 10^6$ M$_{\sun}$) as the
rough dividing line between GCs and UCDs in Fornax.  Nuclear regions of Fornax
dE,Ns were found to be about 0.6 dex more metal poor than Fornax UCDs,
implying that Fornax UCDs do probably not originate from the current
population of dE,Ns.

The Centaurus cluster of galaxies ((m-M)=33.28 mag or 43 Mpc, Mieske
et al.~\cite{Mieske05}) is the most nearby southern cluster of mass
comparable to Virgo (m $\simeq 5\ \times \ 10^{14} M_{\sun}$, Reiprich
\& B\"ohringer~\cite{Reipri02}).  It is therefore a natural target for
broadening the environmental baseline of UCD investigations. The
Centaurus cluster consists of a main component Cen30 at $v_{\rm
  rad}\simeq$ 3000 km/s, dominated by NGC 4696, and an in-falling,
spiral-rich subcomponent at $v_{\rm rad}\simeq$ 4500 km/s, called
Cen45 (Stein et al.~\cite{Stein97}), which is dominated by NGC 4709.

In this paper, we present a spectroscopic search for UCDs in
the Centaurus cluster, which we term Centaurus Compact Object Survey ``CCOS''.

\begin{figure*}
\begin{center}
  \epsfig{figure=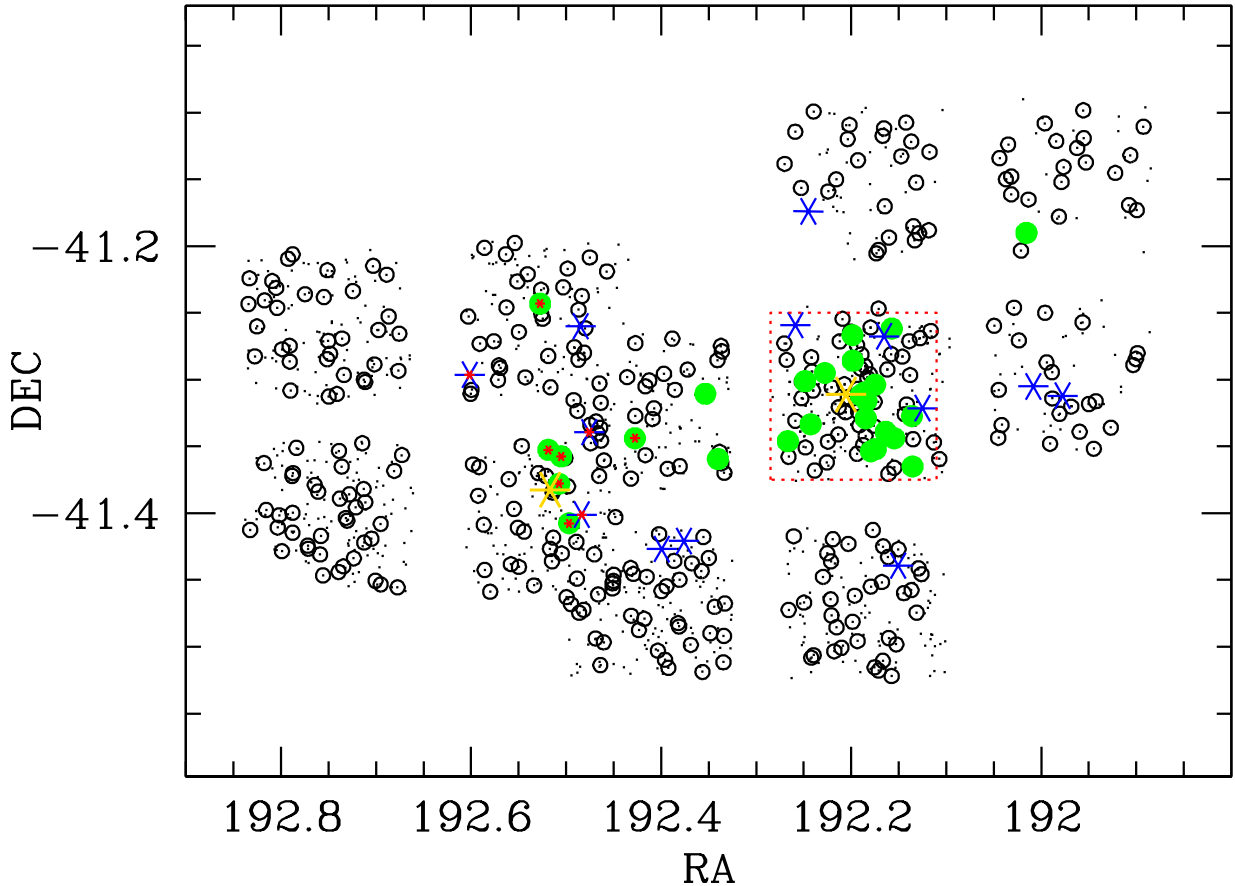,width=17.2cm}
       \caption{Map of the VIMOS observations in the Centaurus cluster. Dots are all photometrically selected compact object candidates from the VIMOS pre-images. Open circles mark the 386 compact objects with measured radial velocities (out of 405 for which slits had been allocated). Filled (green) circles indicate the 27 compact objects with radial velocities in the range of the Centaurus cluster ($1750<v_{rad}<5550$ km/s, see Fig.~\ref{vradhistall}). Filled (green) circles with red dots indicate objects with radial velocity above 4000 km/s, hence members of the Cen45 sub-cluster. Blue asterisks indicate the locations of 13 dwarf elliptical galaxies included in this study. The golden asterisks indicate the location of NGC 4696 and NGC 4709, the central galaxies of the Cen30 and Cen45 components. The dotted red rectangle indicates the central FORS pointing of our deep photometric Centaurus cluster study (see text and Mieske et al.~\cite{Mieske05}).}  
\label{map}
\end{center}
\end{figure*}

\section{The data}
\label{data}
The data for this publication were obtained in service mode with the
VIsible MultiObject Spectrograph VIMOS (Le Fevre et
al.~\cite{Lefevr03}) mounted on UT3 Melipal at the VLT (programme
076.B-0293). VIMOS allows simultaneous observing of 4 quadrants, each
of dimension $7'\times 8'$, and separated by about 2$'$. We observed three multi-object spectroscopy (MOS) pointings in the
central Centaurus cluster, covering both NGC 4696 and NGC 4709 (see
Fig.~\ref{map}).

\subsection{Candidate selection}
\label{candsel}
The candidates for our survey were selected from the VIMOS
pre-imaging that was performed in the $V$ and $R$ filters.  

For de-redenning the apparent magnitudes we used Schlegel et
al.~(\cite{Schleg98}).  To select sources as compact object
candidates, we defined three criteria regarding size, colour and
luminosity. 

1. Be unresolved on the VIMOS pre-imaging (as judged by SExtractor
star-classifier, Bertin \& Arnouts~\cite{Bertin96}). The PSF FWHM
typically was 0.8$''$, corresponding to $\approx$ 150 pc at the
distance of Centaurus.

2. Have de-redenned colours $0.23<(V-R)_0<0.73$ mag, in order to cover
a broad metallicity range for old stellar populations ($-3\le{\rm [Fe/H]}\le 0.5$ dex for a 13 Gyr population, Bruzual \& Charlot~\cite{Bruzua03}).

3. Have de-redenned apparent magnitudes $19.2<V_0<22.4$ mag
($-14.1<M_V<-10.9$ mag). This covers both the bright end of the
globular cluster luminosity function (GCLF) and the magnitude range of
all UCDs discovered so far.

\subsection{Spectroscopic observations}
Within our 12 masks (3 pointings $\times$ 4 quadrants) the VIMOS mask
creation software VMMPS enabled the allocation of 405 compact objects
(minimum slit length 6$''$), compared to a total of 1340
photometrically selected sources.  We were able to measure redshifts
for 386 out of those 405 sources.  Our completeness in the area
surveyed is hence about 29\%. We furthermore added 13 early-type dwarf
galaxies in the range $-16.2<M_B<-13.9$ mag to the masks, all from the
Centaurus Cluster Catalog (Jerjen et al.~\cite{Jerjen97}). Nine of
those were already known members from radial velocity measurements
(Stein et al.~\cite{Stein97}).

We used the medium resolution MR grism with the order sorting filter
GG475. This covers the wavelength range from 4800 to 10000  {\AA} at a
dispersion of 2.5 {\AA} per pixel. The average seeing for the
spectroscopic observations was around 0.8$''$, at a slit width of
1.0$''$.  With a pixel scale of 0.2$''$, the instrumental resolution
(FWHM) is 10-12 \AA, at about the limit for Lick index measurements.
For each pointing the total exposure time was 8400 seconds, subdivided
in four single exposures of 2100 seconds.

Arc-lamp exposures for wavelength calibration were taken for every second
science exposure.  We also observed four Lick
standard stars (HD064606, HD131976, HD131977, HD148816) with the same
grism and slit width as the science exposures.
\section{Data reduction}
\label{reduction}
For the data reduction from 2D raw spectrum to wavelength calibrated
1D spectrum we used the recipes vmmosobsjitter and vmmosobsstare
provided by the ESO VIMOS
pipeline\footnote{http://www.eso.org/projects/dfs/dfs-shared/web/vimos/vimos-pipe-recipes.html}.
These recipes perform bias subtraction, flat field division,
wavelength calibration, image combination (only for vmmosobsjitter),
and spectrum extraction.

The radial velocity measurements of the calibrated 1D spectra were
performed via cross-correlation using the IRAF task fxcor in the RV
package.  As template for cross-correlation we used a synthetic
spectrum created to resemble a typical early-type galaxy (see also
Mieske et al.~\cite{Mieske02}).

\begin{figure}
\begin{center}
  \epsfig{figure=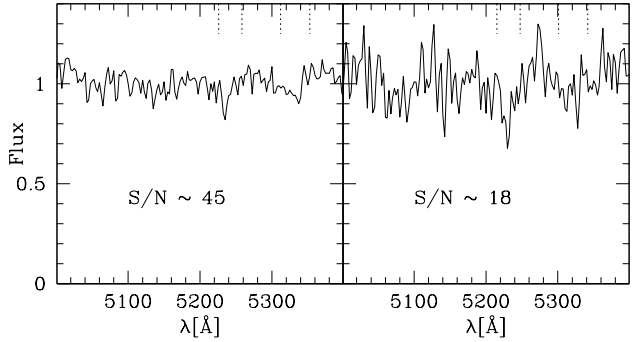,width=8.6cm}
\caption{Example excerpts of two reduced spectra of Centaurus compact objects. The dashed ticks at the top mark the correspondingly red-shifted regions of the Mgb and Fe5270 Lick indices. {\bf Left:} Object CCOS J1248.54-4119.64, with a S/N around 45. {\bf Right:} Object CCOS J1248.74-4118.95, with a S/N around 18.}  
\label{spectra}
\end{center}
\end{figure}

\begin{figure}[h!]
\begin{center}
  \epsfig{figure=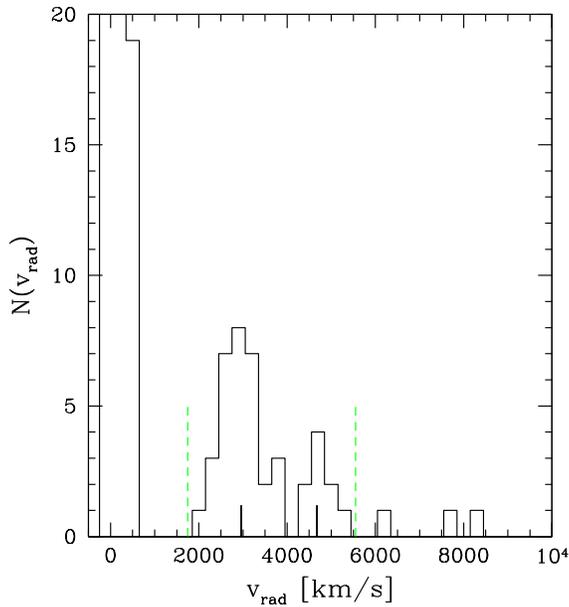,width=8.6cm}
\caption{Radial velocity histogram of all observed sources. The green dashed lines mark the limits applied for Centaurus cluster membership selection (40 Centaurus cluster members, including 13 dEs). The two vertical ticks mark the radial velocities of NGC 4696 ($\sim 3000$ km~s$^{-1}$) and NGC 4709 ($\sim 4500$ km~s$^{-1}$).}  
\label{vradhistall}
\end{center}
\end{figure}

For the cluster membership determination we measured the radial
velocity on the 1D spectrum combined from the four single exposures
with the recipe vmmosobsjitter. The S/N per pixel in these spectra was
in the range 12 to 63 in the wavelength range of highest transmission
(6700 to 6800 \AA).  The radial velocity measurement errors were of
the order 50-100 km/s.  As a cluster membership criterion we required
$1750<v_{rad}<5550$ km/s, excluding both foreground stars and
background galaxies (see Fig.~\ref{vradhistall}). This resulted in the
discovery of 40 cluster members, comprised by 27 compact cluster
members and the 13 dEs (see Tables~\ref{table} and~\ref{table_dE}).
Example spectra for two Centaurus compact objects are shown in
Fig.~\ref{spectra}.

In order to refine the velocity measurements, we compared the radial
velocities derived from the combined exposures with those derived from
the single science exposures that were observed back-to-back with an
arc-lamp exposure. This comparison showed no systematic difference in
quadrants 3 and 4 to a level of 30 km/s.  For quadrants 1 and 2 the
velocity derived from the two exposures next to the wave-lamp was
higher by about 100 km/s for quadrant 1 and lower by about 100 km/s
for quadrant 2. We attribute these offsets to the well-known flexure
of the VIMOS instrument. In order to determine the final radial
velocity, we corrected the mean velocity in quadrants 1 and 2 by those
shifts, and adopted the uncorrected mean for quadrants 3 and 4.

\begin{figure}
\begin{center}
  \epsfig{figure=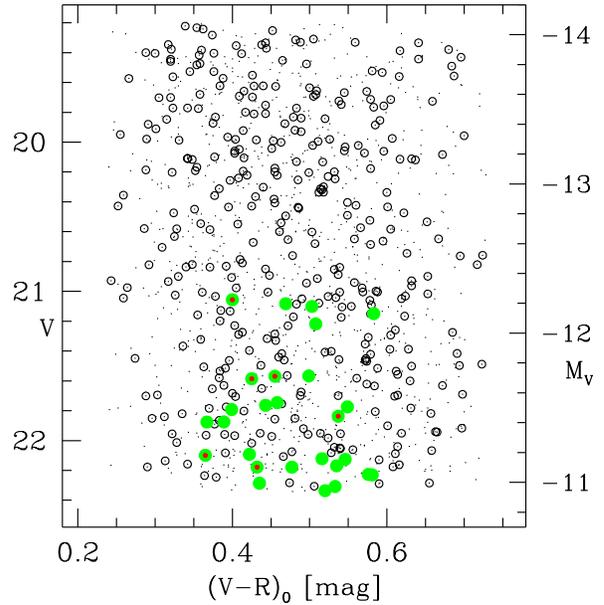,width=8.6cm}
       \caption{Colour-magnitude diagram in VR of the sources from Fig.~\ref{map}, excluding the dwarf elliptical galaxies. }  
\label{cmd}
\end{center}
\end{figure}
\begin{figure}[]
\begin{center}
  \epsfig{figure=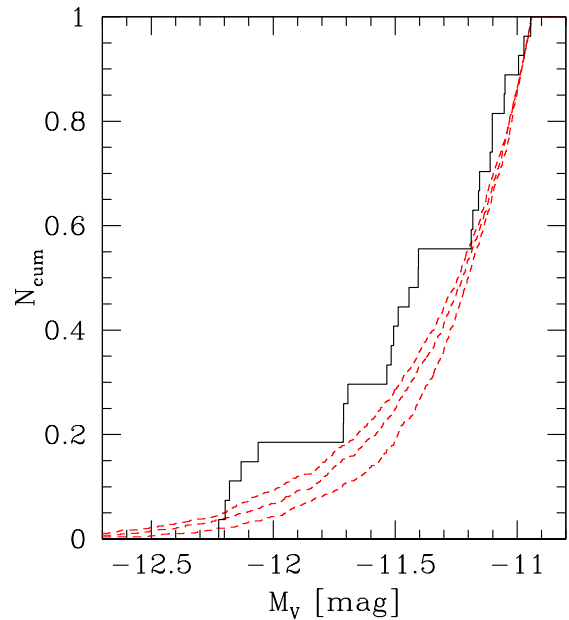,width=8.6cm}
\caption{Cumulative magnitude distribution of Centaurus compact objects. Dashed lines indicate the cumulative magnitude distribution of assumed Gaussian Globular Cluster Luminosity Functions with three different widths $\sigma$: from top to bottom 1.4, 1.3, and 1.2 mag. The assumed turnover magnitude is $M_V=-7.4$ mag. The KS-test probabilities with which the measured compact object distribution is drawn from the Gaussians are 21\% ($\sigma=1.4$ mag), 11\% ($\sigma=1.3$ mag) and 2\% ($\sigma=1.2$ mag).}  
\label{vhist}
\end{center}
\end{figure}

\section{Results}
\label{results}
Out of the 386 compact objects with measurable redshift observed, 27
turn out to be Centaurus cluster members (see Fig.~\ref{vradhistall}
and Table~\ref{table}). In the following subsections we will
investigate their photometric properties, spatial distribution,
abundances, kinematics, and their structural parameters. The null
hypothesis is that their properties can be explained by the Centaurus
cluster globular cluster systems.

\subsection{Photometry}
\label{photometry}
Fig.~\ref{cmd} shows a colour-magnitude diagram in $VR$ of all compact
cluster members, all observed objects, and all photometrically
selected objects. It suggests that no detection bias was introduced
by our colour selection.

A first step to test our null hypothesis is to compare the cumulative
magnitude distribution of the compact objects with that expected for a
generic globular cluster luminosity function (GCLF).  For very bright
galaxies like NGC 4696 and NGC 4709, the typical dispersion ($\sigma$)
of a Gaussian GCLF is around 1.3 to 1.4 mag (Jordan et
al.~\cite{Jordan06}~and~\cite{Jordan07}, Kundu \&
Whitmore~\cite{Kundu01}).  Fig.~\ref{vhist} shows that for such
values, the agreement with our data is still at the 10-20\% level
according to a KS-test.  Only with respect to an assumed width of
$\sigma=1.2$ mag, the agreement drops to 2\%.

A further test is to estimate the total number of UCDs one would
expect to detect in the case of the stripping scenario. Bekki et
al.~(\cite{Bekki03}) predict about two dozen stripped nucleated dwarfs
in the range $-13<M_B<-11$ mag in the central 200kpc projected
distance of the Virgo cluster. A comparable number may be expected for
the Centaurus cluster given its similar mass.  The completeness of our
survey within 200 kpc (16$'$) projected radius is about 12\%,
induced by 29\% slit allocation completeness and about 40\% area coverage.
Formally, we would thus expect 3$\pm 2$ stripped dwarf galaxies with
$M_B<-11$ mag ($M_V<-11.5$ mag). This number is too small to support
the hypothesis that the slight overpopulation seen in Fig.~\ref{vhist}
for bright luminosities is caused by stripped dwarfs.

Finally, we compare the total number of 27 compact objects with that
expected from the GCLF. Our faint magnitude cut of $M_V=-10.9$ mag
restricts us to 0.5 $^{+0.35}_{-0.2}$\% of all GCs, provided that
their luminosity distribution is described by a Gaussian of width
$\sigma=1.3 \pm 0.1$ mag and turn-over magnitude $M_V=-7.4 \pm 0.2$
mag. Under the null hypothesis that our compact objects are all
explained by the GC system, we would deduce a total number of GCs in
the surveyed area of $\frac{27}{0.005\times
  C}=\frac{5400}{C}=\frac{5400}{0.28}=19300^{+12900}_{-8260}$, where
$C=0.28$ is the completeness of our survey in the magnitude range
where compact objects are discovered. We can now estimate which
specific GC frequency $S_N=N_{GC}\times 10^{0.4\times (15+M_V)}$ is
required for the two main Centaurus galaxies NGC 4696 and NGC 4709 in
order to produce a total amount of $\sim$20000 GCs. We derive the
total absolute magnitude of both galaxies from our $V$-band pre-images.
Under the assumption of $(m-M)=33.28$ mag, we obtain values of
$M_V=-23.1$ mag for NGC 4696 and $M_V=-22.3$ mag for NGC 4709, in
agreement with other literature estimates (e.g. Michard et
al.~\cite{Michar05}, Jerjen et al.~\cite{Jerjen97}). In order to
contain 19300$^{+12900}_{-8260}$ GCs, both galaxies would require
$S_N=7.5^{+4.9}_{-2.3}$.  This agrees with the $S_N$ values of 7.3
$\pm$ 1.5 (NGC 4696) and 5.0 $\pm$ 1.3 (NGC 4709) derived in Mieske et
al.~(\cite{Mieske05}) for the central regions of both galaxies (radii
below 2$'$).

We conclude that the luminosity distribution and number of compact
objects are consistent with the GC systems of NGC 4696 and NGC 4709.

\subsection{Spatial distribution}
In this sub-section we compare the cumulative radial distribution of
the compact objects with that of dEs and of GCs.

\begin{figure}[]
\begin{center}
  \epsfig{figure=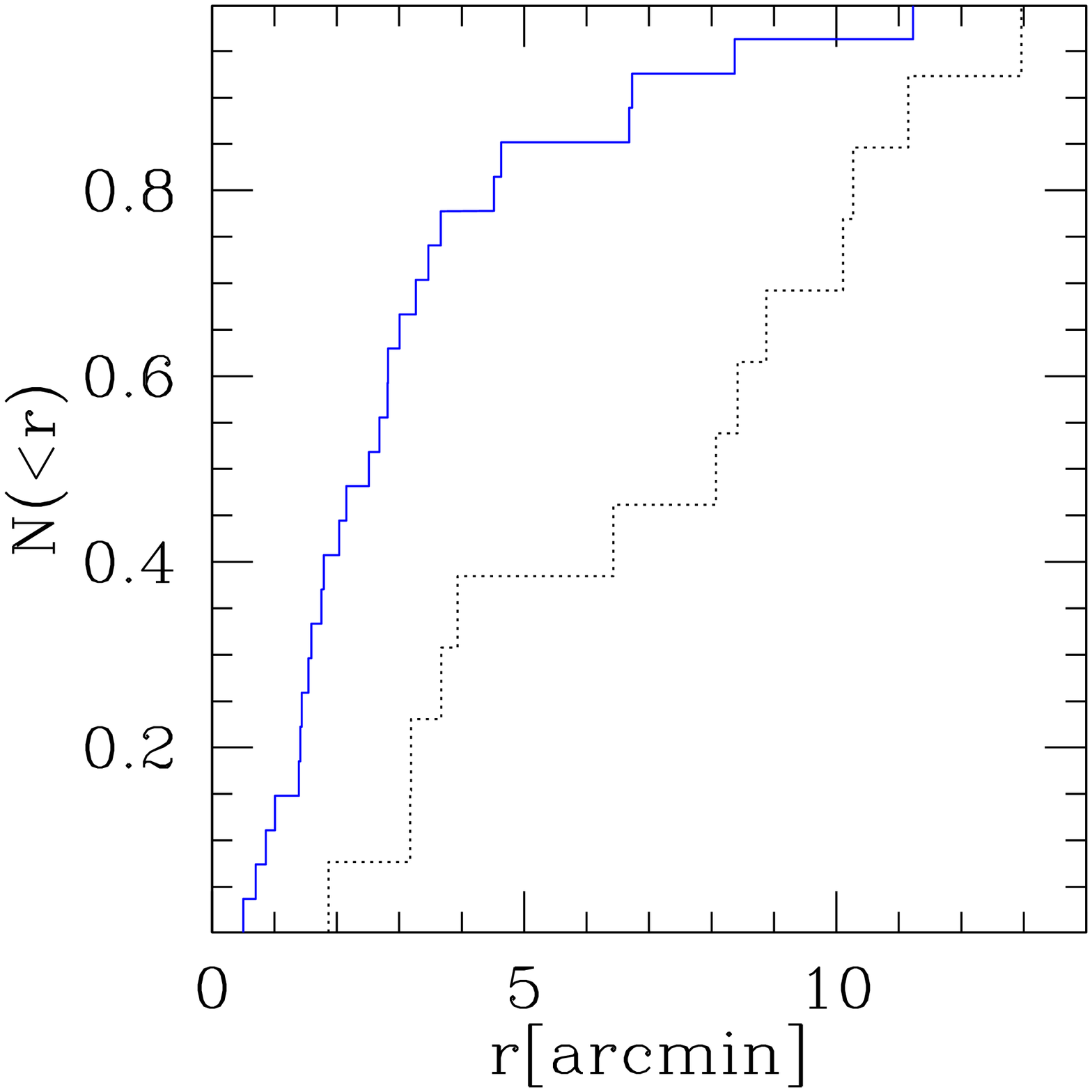,width=4.3cm}
  \epsfig{figure=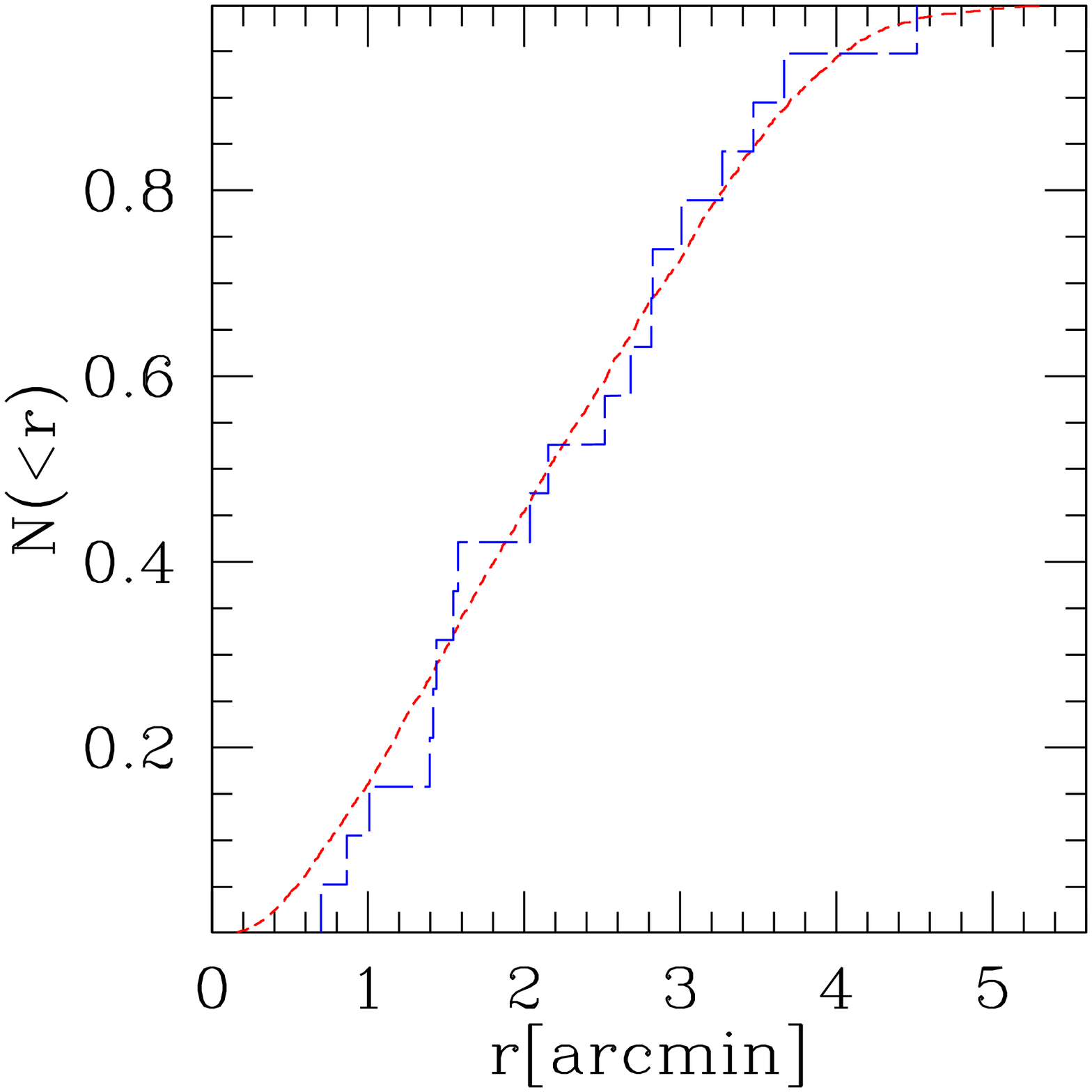,width=4.3cm}
       \caption{Comparison of cumulative radial distributions of various object samples. {\bf Left panel:} Centaurus compact objects are indicated by the (blue) solid histogram. The sample of 13 dwarf ellipticals is indicated by the (black) dotted histogram. For this plot, the radial distance was calculated with respect to NGC 4696 for the Cen30 members and with respect to NGC 4709 for the Cen45 members. The samples of compact objects and dEs have identical parent distributions at the 0.2\% confidence level according to a KS-test. {\bf Right panel:} The short dashed red line is the distribution of GC candidates around NGC 4696 from Mieske et al.~(\cite{Mieske05}), see Fig.~\ref{map}. The long dashed blue histogram is the distribution of compact objects restricted to the same field as the GC candidate sample. According to a KS test, both samples stem at 98\% probability from the same parent distribution.}  
\label{radcum}
\end{center}
\end{figure}

Fig.~\ref{radcum} shows that the compact objects are clearly more
concentrated towards the central cluster galaxies than are the dEs
included in our study. Selection effects regarding the dE sample
should not significantly influence this finding, given that the dEs
were selected from a homogeneous source catalog (Jerjen et
al.~\cite{Jerjen97}). Note that the stronger clustering of the compact
objects does not exclude that they originate from dEs, since tidal
stripping is expected to preferably affect dEs with smaller
apocentric distances (Bekki et al.~\cite{Bekki03}).

For comparison with the genuine GC population, we use the deep FORS
$VI$ photometry from Mieske et al.~(\cite{Mieske05}). 
The central FORS pointing of our previous study practically matches
the area covered by the VIMOS quadrant centred on NGC 4696 (see
Fig.~\ref{map}). Since most of the candidates were detected in that
quadrant, a direct comparison is possible. To define the GC candidate
sample in the FORS observations, we restricted the colour range to
$0.7<(V-I)<1.4$ mag (see Sect.~\ref{candsel}) and the magnitude range
to $-7.9>M_V>-10.9$ mag.  The bright magnitude cut makes the GC
candidate sample disjunct from the compact object sample, the faint
magnitude cut is due to the completeness limit. Fig.~\ref{radcum}
shows that the radial distribution of compact objects agrees very well
with that of the GC candidates.  The KS test gives a probability of
98\% that both samples have identical parent distributions.

We thus conclude that the radial distribution of compact objects is
consistent with that of the globular cluster system. 
\begin{figure}[h!]
\begin{center}
  \epsfig{figure=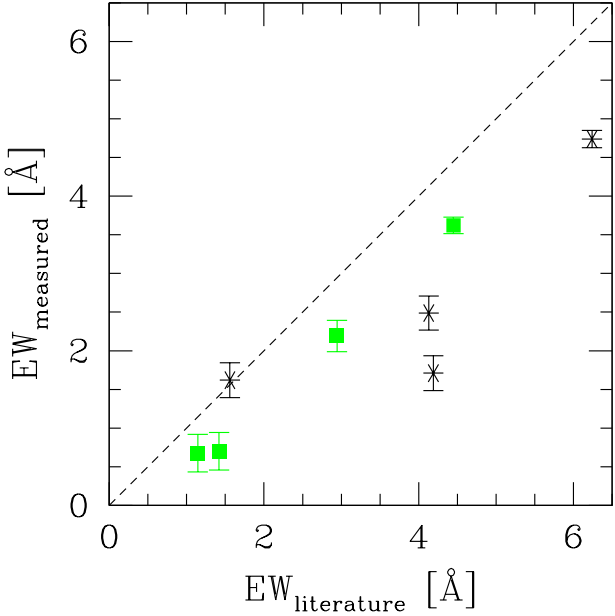,width=8.6cm}
\caption{Comparison between  measured and literature Lick index values for the four Lick standard stars included in our observations. Index ${\rm <Fe\rangle }$ is indicated by green solid squares, Mgb by black asterisks. Literature data is from Worthey et al.~(\cite{Worthe94}). The dashed line indicates the identity. }  
\label{licks}
\end{center}
\end{figure}

\subsection{Abundances}
\label{met}
Line index measurements of the 27 confirmed Centaurus compact objects,
the 13 dEs, and the four Lick standard stars were performed with
standard IRAF routines within the ONEDSPEC package.  We derived Lick
indices using the pass-band definitions of Trager et
al.~(\cite{Trager98}). The instrumental resolution of  10-12 {\AA}
is slightly worse than the Lick resolution of 8-9 \AA.  The
pass-bands were red-shifted according to the radial velocity of each
investigated object.  

For the four Lick standard stars, we compare in Fig.~\ref{licks} our
equivalent width (EW) measurements of both Mgb and $\langle {\rm Fe}
\rangle$ with the literature values (Worthey et al.~\cite{Worthe94}).
We determine a global offset of $-$0.69 $\pm$ 0.08 {\AA} for $\langle
{\rm Fe}\rangle $ and $-$1.39 $\pm$ 0.53 {\AA} for Mgb.  The
correction is quite uncertain for Mgb, mainly due to one outlier
(HD148816). We have gone back to the raw spectrum, but did not find
artefacts that would have inhibited a correct index measurement for
this source. Furthermore, the S/N was sufficiently high ($\ge 50$) for
all four Lick standards. We re-measured the line indices for continuum
normalised spectra, for logarithmic re-binnings, and for spectra
extracted with a task outside the VIMOS pipeline (apall in the IRAF
package twodspec). However, all this did not change the index values
by more than $\sim$ 0.2 \AA, and did not remove the outlier. We
therefore apply the offsets resulting from the mean of all four Lick
standards to the measured indices of compact objects and dEs.

\begin{figure}[h!]
\begin{center}
  \epsfig{figure=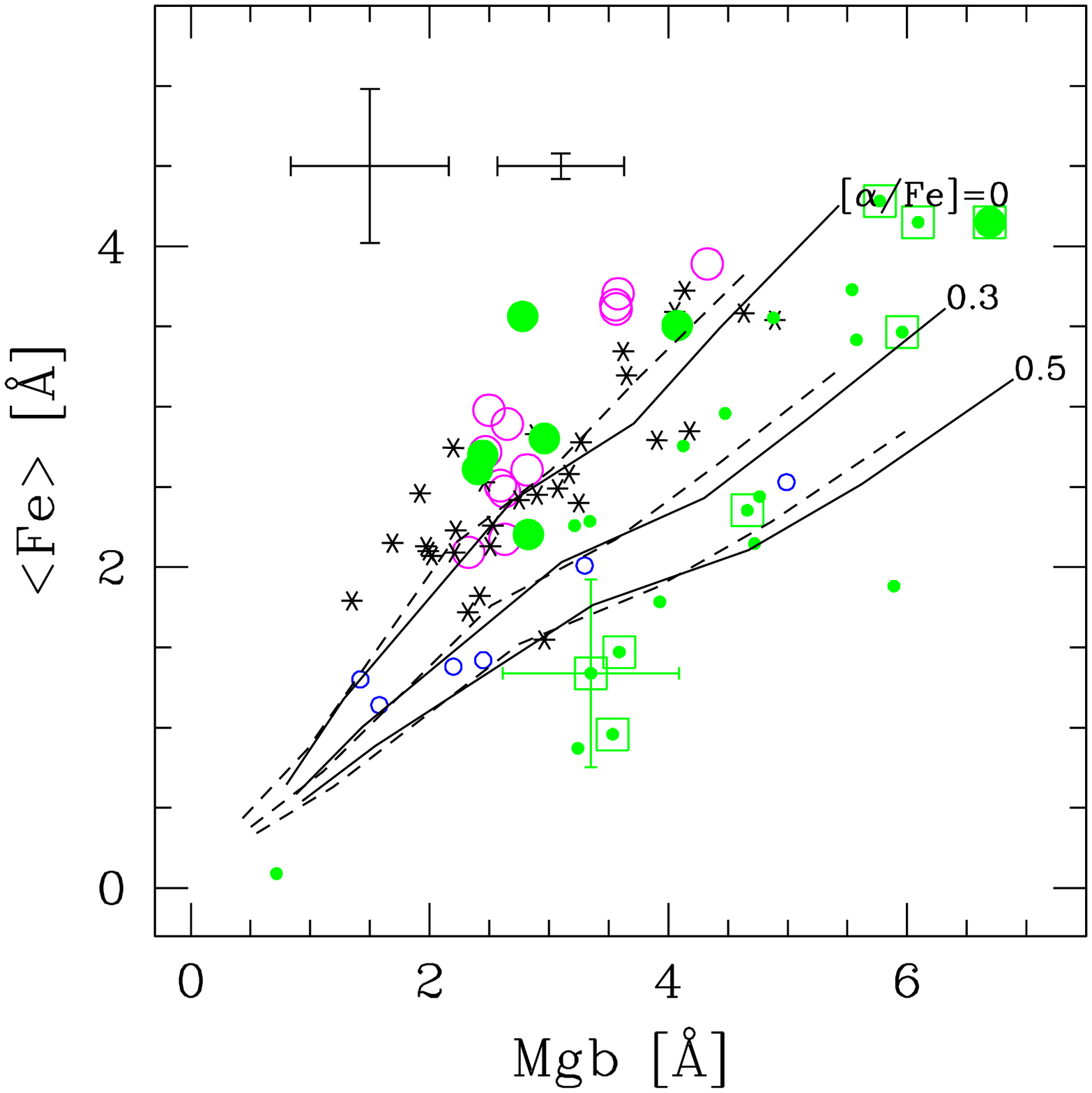,width=8.6cm}
  \caption{Lick index ${\rm \langle Fe\rangle }$ plotted vs. Mgb. Large, green filled circles are Centaurus compact objects with S/N$>$30, small green filled circles those with S/N$<$30. At the top, the left error bar indicates the average statistical uncertainty of our (green) data points, while the right error bar indicates the global uncertainty of our Lick calibration (see text). Sources surrounded by a solid square indicate those with HST imaging available (see Sect.~\ref{struct}). The source with an attached error bar is the most extended of those compact objects (see Sect.~\ref{struct}). Asterisks indicate values for central regions of dwarf elliptical galaxies, comprising the 13 Centaurus cluster dEs included in our study, and Fornax and Virgo dEs from Geha et al.~(\cite{Geha03}). Blue open circles are Virgo UCDs from Evstigneeva et al.~(\cite{Evstig07}). Magenta open circles are the Fornax UCDs (Mieske et al.~\cite{Mieske06a}). The solid (dashed) lines indicate SSP models by Thomas, Maraston \& Bender~(\cite{Thomas03}) for an assumed age of 12 Gyrs (4 Gyrs) and the three indicated [$\alpha$/Fe] values. Metallicities are from bottom to top -2.25 to 0.67 dex. 
}  
\label{Fe_V1}
\end{center}
\end{figure}

\begin{figure*}[]
\begin{center}
  \epsfig{figure=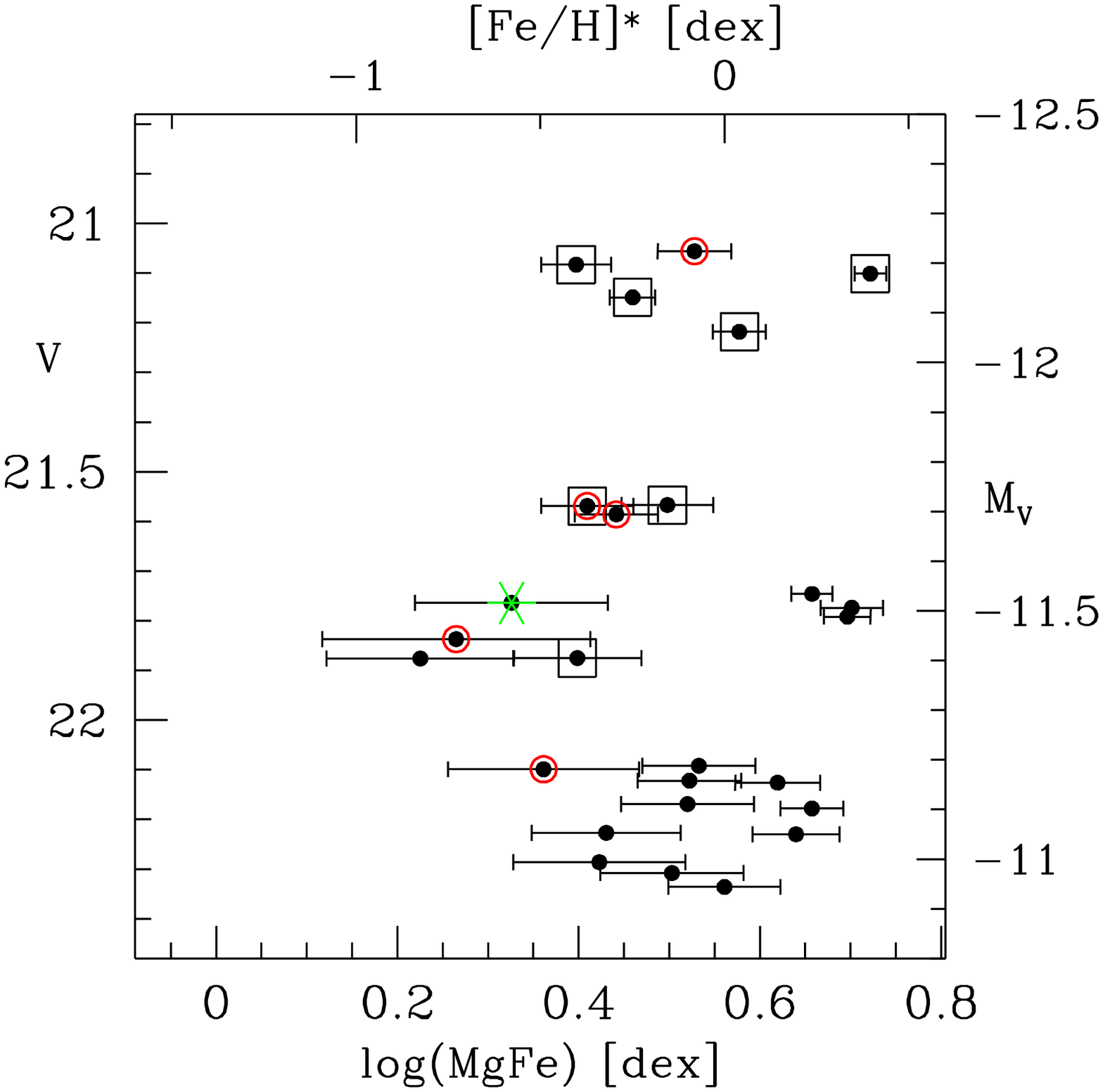,width=8.6cm}
  \epsfig{figure=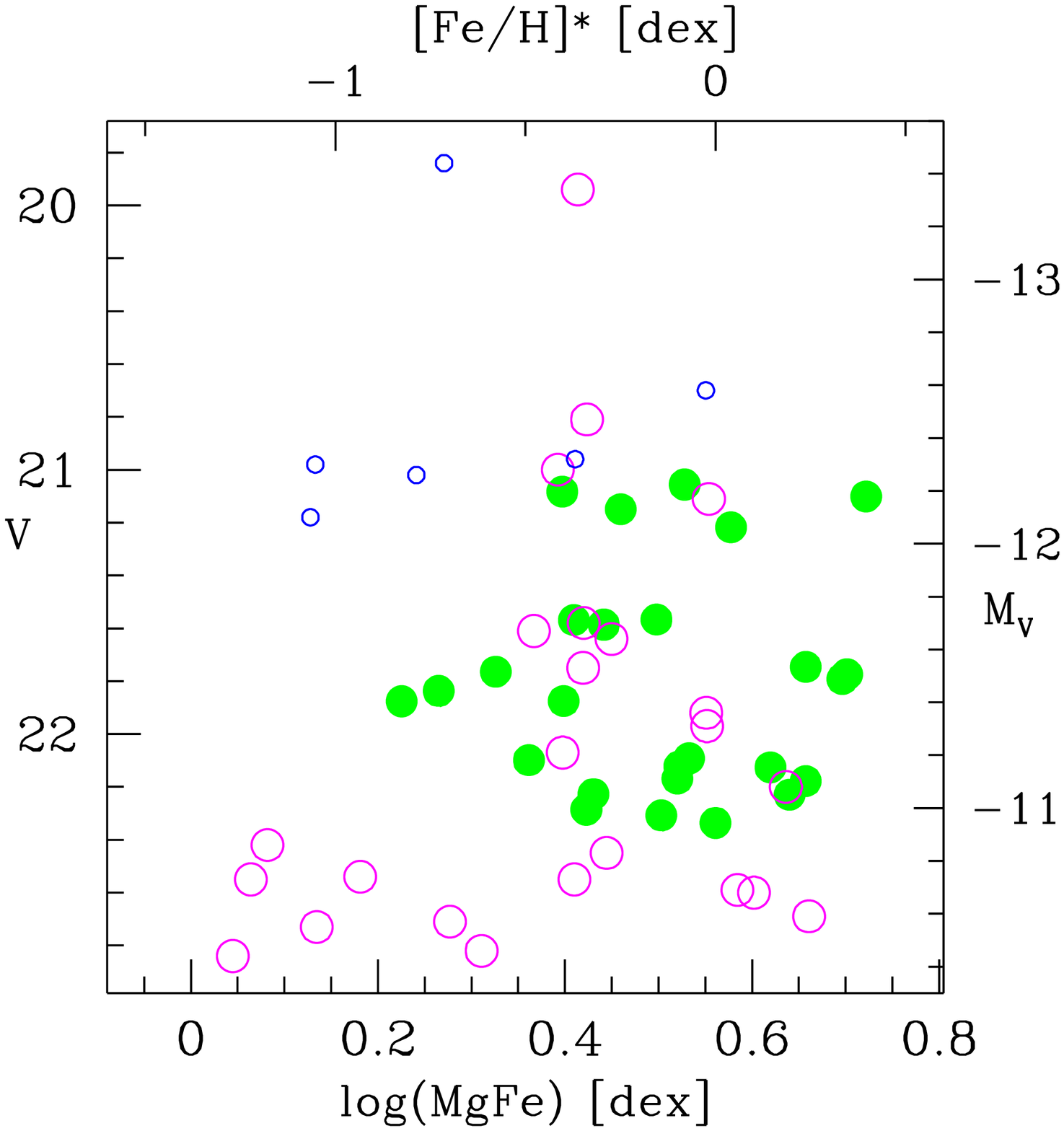,width=8.6cm}
\caption{{\bf Left panel:} Logarithm of the MgFe line index plotted vs. V for the Centaurus compact objects. The upper x-axis indicates an estimate of the corresponding [Fe/H], accurate to 0.2-0.3 dex, see text. Symbols surrounded by (red) circles indicate Cen45 members. The solid squares indicate spectra with S/N larger than 30. The green asterisk indicates the most extended object among the 8 sources with HST imaging (see text). {\bf Right panel:} Log(MgFe) plotted vs. V for three samples of compact objects. Green filled circles indicate the Centaurus compact objects from the left panel. Blue open circles are the Virgo UCDs (Evstigneeva et al.~\cite{Evstig07}). Magenta open circles indicate Fornax compact objects (Mieske et al.~\cite{Mieske06a}).}  
\label{Fe_V2}
\end{center}
\end{figure*}

In Fig.~\ref{Fe_V1} we plot the correspondingly corrected ${\rm
  \langle {\rm Fe}\rangle }$ and Mgb line index EWs against each
other. In the plot we also indicate the global uncertainty of the
  Lick calibration for $\langle {\rm Fe}\rangle $ and Mgb. We
sub-divide the Centaurus compact object sample into a high and low S/N
sub-sample using S/N=30 as limit. This roughly corresponds to a
magnitude limit of $M_V=-11.6$ mag, see Fig.~\ref{Fe_V2}.  We
furthermore show literature line index measurements for Fornax and
Virgo UCDs (Mieske et al.~\cite{Mieske06a}, Evstigneeva et
al.~\cite{Evstig07}), Centaurus dEs (this paper), and Fornax and Virgo
dEs (Geha et al.~\cite{Geha03}). We overlay three model grids from
Thomas, Maraston \& Bender~(\cite{Thomas03}), corresponding to a 12
Gyr single stellar populations with [$\alpha$/Fe]=0, 0.3, and 0.5 dex.
The UCDs in Fornax, the Centaurus compact objects with high S/N, and
the dEs in all three environments show roughly solar $\alpha$
abundances.  The UCDs investigated in Virgo show in contrast
super-solar abundances, between 0.3 and 0.5 dex. Assuming the IMFs are
roughly similar, this implies that Virgo UCDs have had a more
truncated star formation history than the other samples, in line with
the galactic bulge population of globular clusters (e.g. Puzia et
al.~\cite{Puzia02}, Barbuy et al.~\cite{Barbuy99}, McWilliam \&
Rich~\cite{McWill94}, Carretta et
al.~\cite{Carret01}~and~\cite{Carret07}, Gratton et
al.~\cite{Gratto06}). 

It is interesting to note that also the low S/N Centaurus compact
objects appear to show super-solar $\alpha$ values, albeit at a
considerable scatter.  We have tested whether the corresponding low
$\langle {\rm Fe}\rangle $ values may be artifacts of the lower S/N.
For that we combined rest-frame corrected spectra to a higher S/N
master spectrum, and compared the line index of the combined spectrum
to the mean of the single indices. We did this for 4 sets of 3 spectra
in different S/N ranges from S/N$\sim$18 to S/N$\sim$45. We found only
very small differences of order 0.05~$\AA$ between the combined index
and the averaged index, indicating that lower S/N may not be the
reason for the apparent difference in $\alpha$ values to the brighter
Centaurus compact object. However, deeper data with S/N $\ge 30$
(requiring on-source integration of about 1 night) are certainly
required to confirm the possible trend of $\alpha$ abundance with
luminosity.

In Fig.~\ref{Fe_V2} we plot the logarithm of the [MgFe] index (${\rm [MgFe]}=\sqrt{{\rm Mgb} \ \times \ {\rm \langle {\rm Fe}\rangle }}$)
against luminosity for compact objects in Centaurus, Fornax and Virgo.
The [MgFe] index is a good metallicity indicator (e.g. Puzia et
al.~\cite{Puzia02}), given that it is practically insensitive to
changes in $\alpha$ element contents (e.g. Thomas, Maraston \&
Bender~\cite{Thomas03}). We also indicate an estimate of the
corresponding [Fe/H]. For this we use for [Fe/H]$<-$ 0.4 dex the
empirical calibration between [Fe/H] and [MgFe] derived by Puzia et
al.~(\cite{Puzia02}) from galactic GCs. For [Fe/H]$>-$ 0.4 dex we
adopt an extrapolation of the form ${\rm [Fe/H]_*}=a+b \ \times \ 
\log(\rm MgFe)$ that matches the calibration for [Fe/H]$<-$ 0.4 dex.
This is of course only a rough estimate whose systematic uncertainty
is of the order 0.2-0.3 dex.  The resulting mean [Fe/H]$_*$ of the
Centaurus compact objects is $-$0.14 $\pm$ 0.06 dex ($-$0.08 $\pm$
0.07 dex for the dominating Cen30 component, -0.39 $\pm$ 0.10 dex for
Cen45). This is consistent with the typical metallicity of the
metal-rich globular cluster sub-population (see e.g.  Cohen et
al.~\cite{Cohen03}, Peng et al.~\cite{Peng06}). On a relative scale,
the mean metallicity of the Fornax UCDs is indistinguishable from the
Centaurus compact objects, but slightly higher than the metallicity
range of the Virgo UCDs.


\begin{figure}[h!]
\begin{center}
  \epsfig{figure=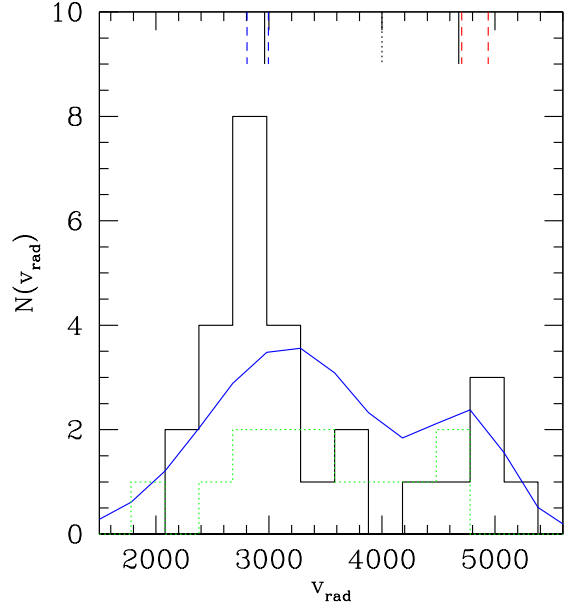,width=8.6cm}
\caption{Radial velocity histogram of Centaurus compact objects (solid line). The (blue) curve indicates the expected velocity distribution if the Cen30 and Cen45 sample had mean velocities and velocity dispersion identical to the early-type galaxy population in both sub-clusters (Stein et al.~\cite{Stein97}). The dotted tick at 4000 km/s indicates the adopted separation between Cen30 and Cen45. The two solid ticks indicate the radial velocities of NGC 4696 (2960 km/s, NED) and NGC 4709 (4680 km/s, NED). The dashed ticks indicate the $\pm$ 1$\sigma$ range of the measured radial velocities for compact objects in Cen30 and Cen45, respectively. The dotted (green) histogram shows the velocity distribution of the 13 dwarf elliptical galaxies included in our sample. }  
\label{vradhist}
\end{center}
\end{figure}
\subsection{Kinematics}
There is no spectroscopic data set of the Globular Cluster System in
Centaurus available. Also, measurements of the field star velocity
dispersion in NGC 4696 or NGC 4709 (Michard et al.~\cite{Michar05},
Blakeslee et al.~\cite{Blakes01}) do not extend to large radii.  The
compact objects' kinematics can therefore only be compared directly
with that of the cluster galaxy population. We sub-divide the sample
of compact objects into Cen30 and Cen45 at a limiting radial velocity
of 4000 km/s (see Fig.~\ref{vradhist}).

Fig.~\ref{vradhist} shows that the mean velocity of the Cen30 and
Cen45 components agree with the velocities of the respective central
galaxies. The velocity dispersion of the 21 Cen30 compact objects is
433 $^{+ 153}_{- 98}$ km/s, where error ranges refer to a 95\%
confidence interval (2$\sigma$). This value is lower at the 4$\sigma$
level than the dispersion of 738 km/s of the Cen30 early-type galaxies
from Stein et al.~(\cite{Stein97}). The velocity dispersion of the six
Cen45 compact objects is 288 $^{+ 315}_{- 95}$ km/s, indistinguishable
to within its large error range from the dispersion of 345 km/s of the
Cen45 early-type galaxies.

The lower velocity dispersion of the Cen30 compact objects as compared
to the cluster galaxy population is consistent with the more centrally
clustered distribution of the former. It indicates that the compact
objects are associated more to the central galaxies than to the
overall cluster potential.
    
\subsection{Structural Parameters}
\label{struct}
\begin{figure}[h]
\begin{center}
  \epsfig{figure=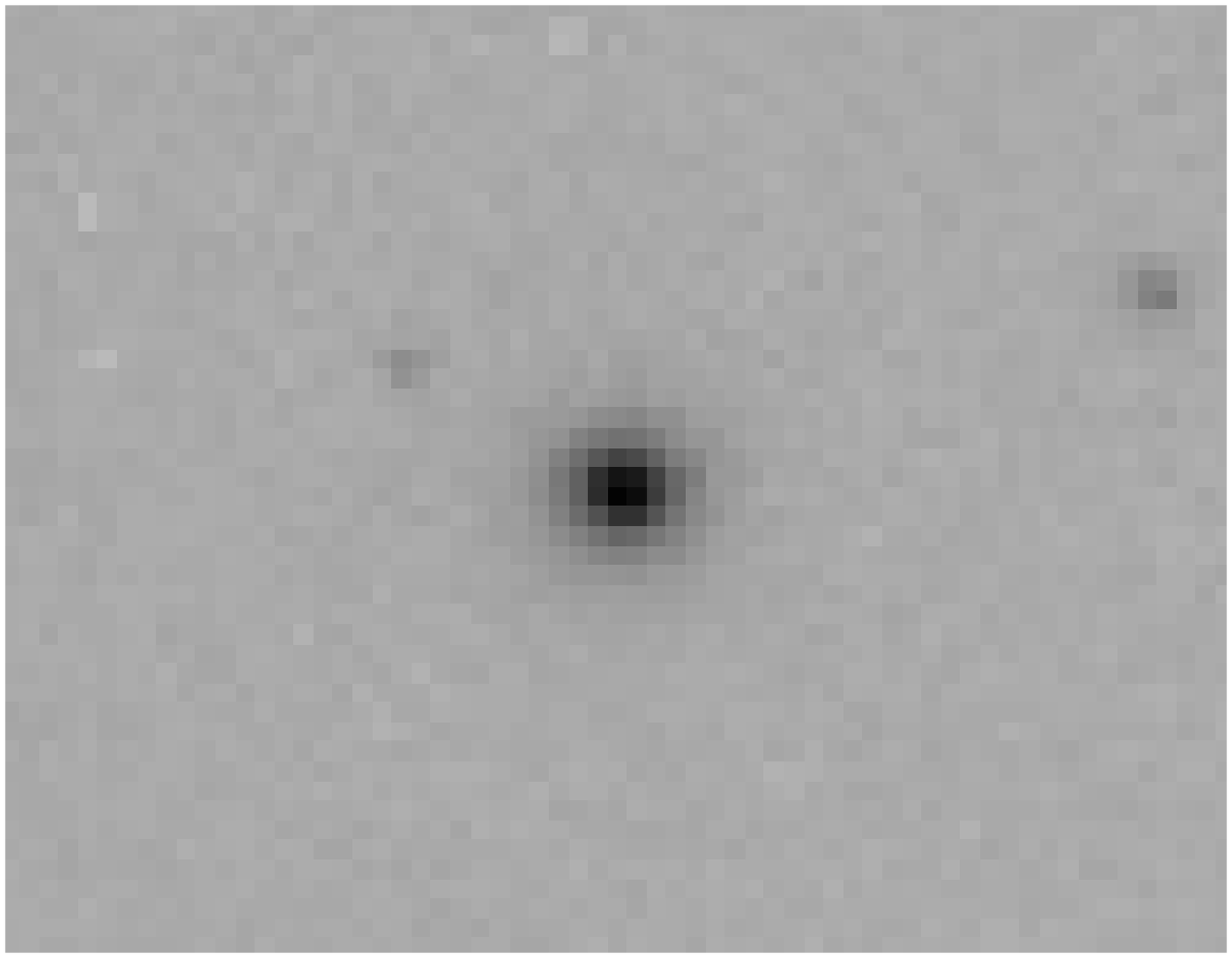,width=4.2cm}\hspace{-4cm}\parbox[t]{4cm}{\vspace{-0.4cm}\hspace{0.1cm}\bf   CCOS J1248.74-4118.58}\hspace{0.2cm}
  \epsfig{figure=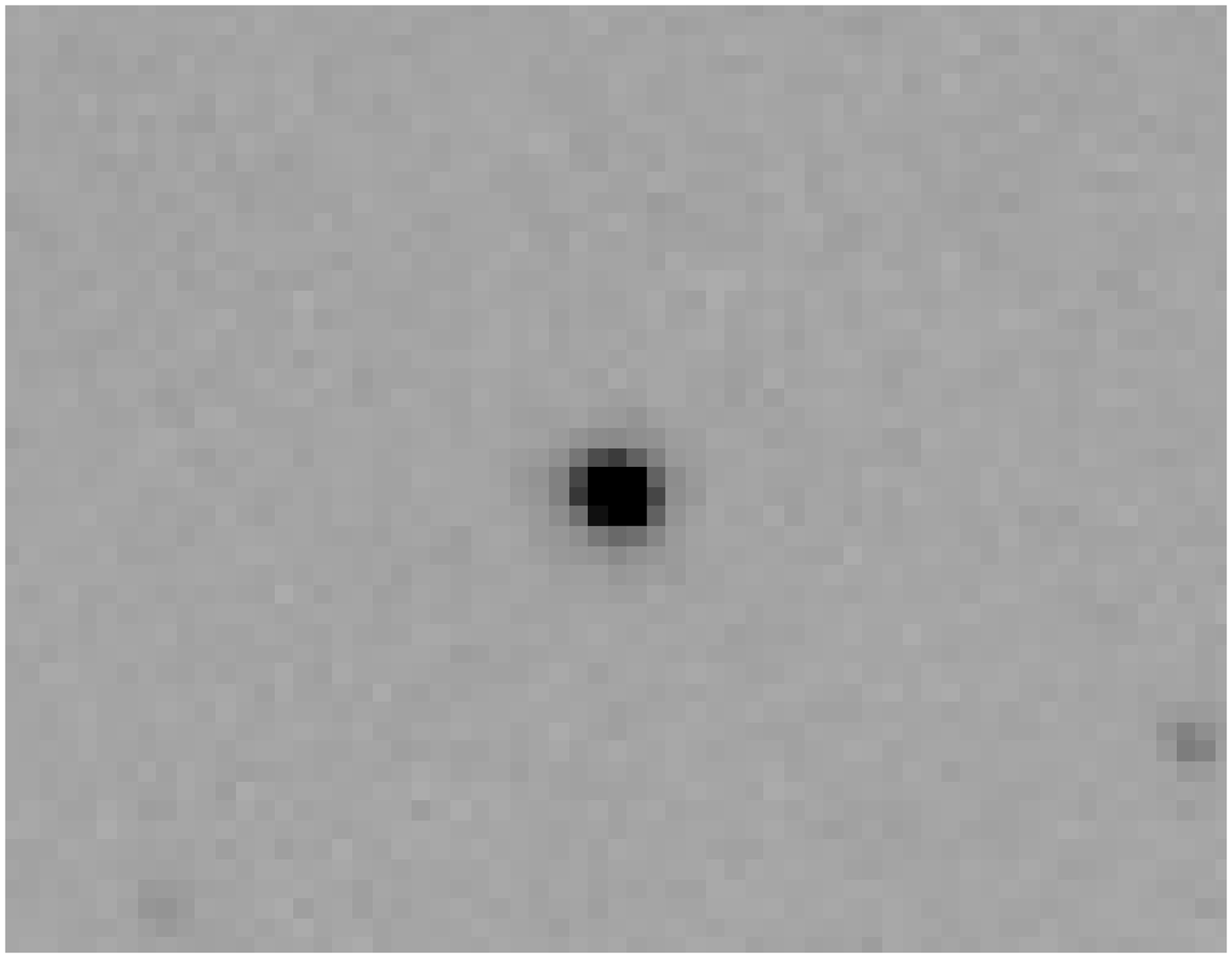,width=4.2cm}\hspace{-4cm}\parbox[t]{4cm}{\vspace{-0.4cm}\hspace{0.1cm}\bf   CCOS J1248.76-4118.70}\\
\epsfig{figure=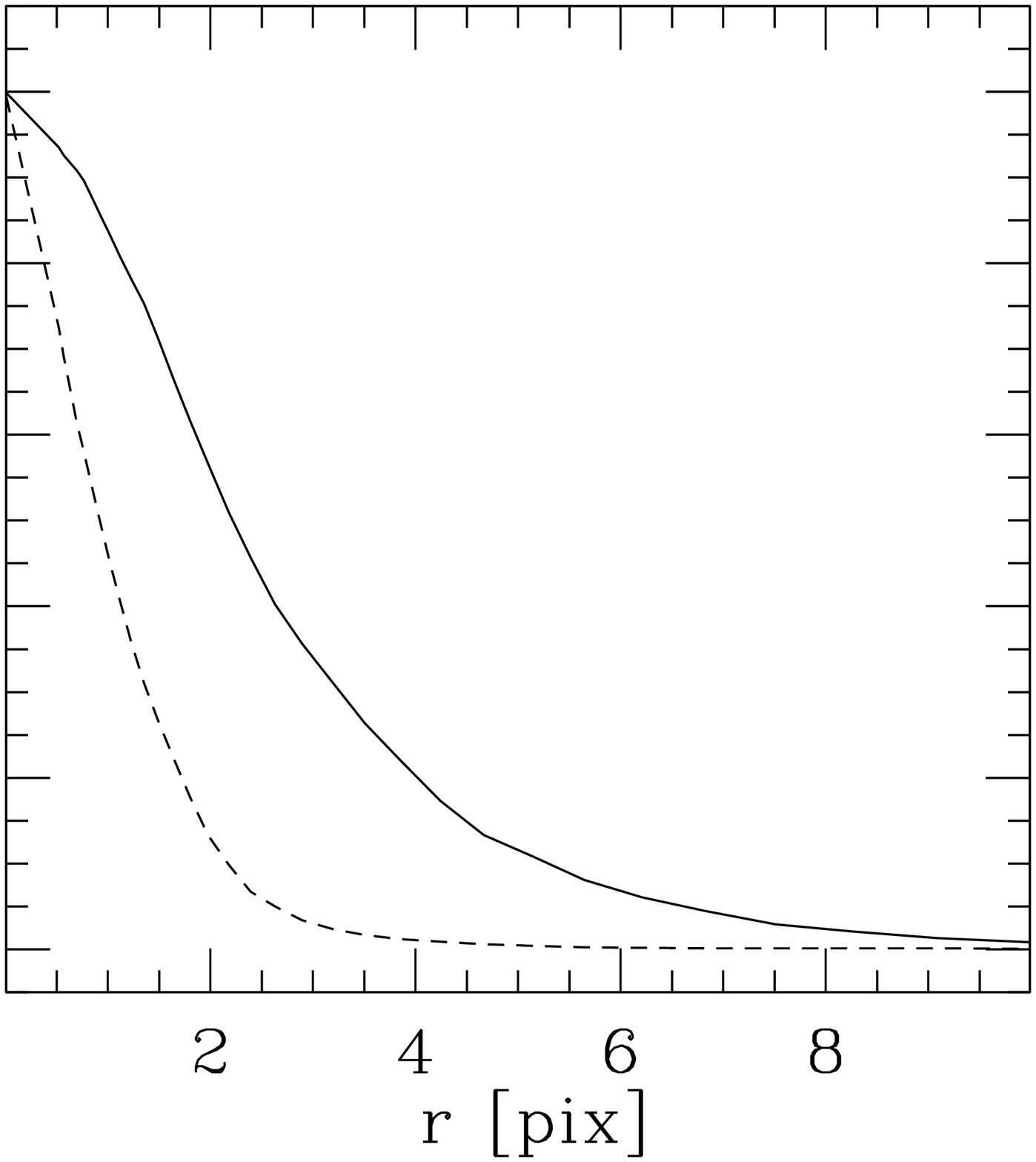,width=4.2cm}\hspace{0.2cm}
\epsfig{figure=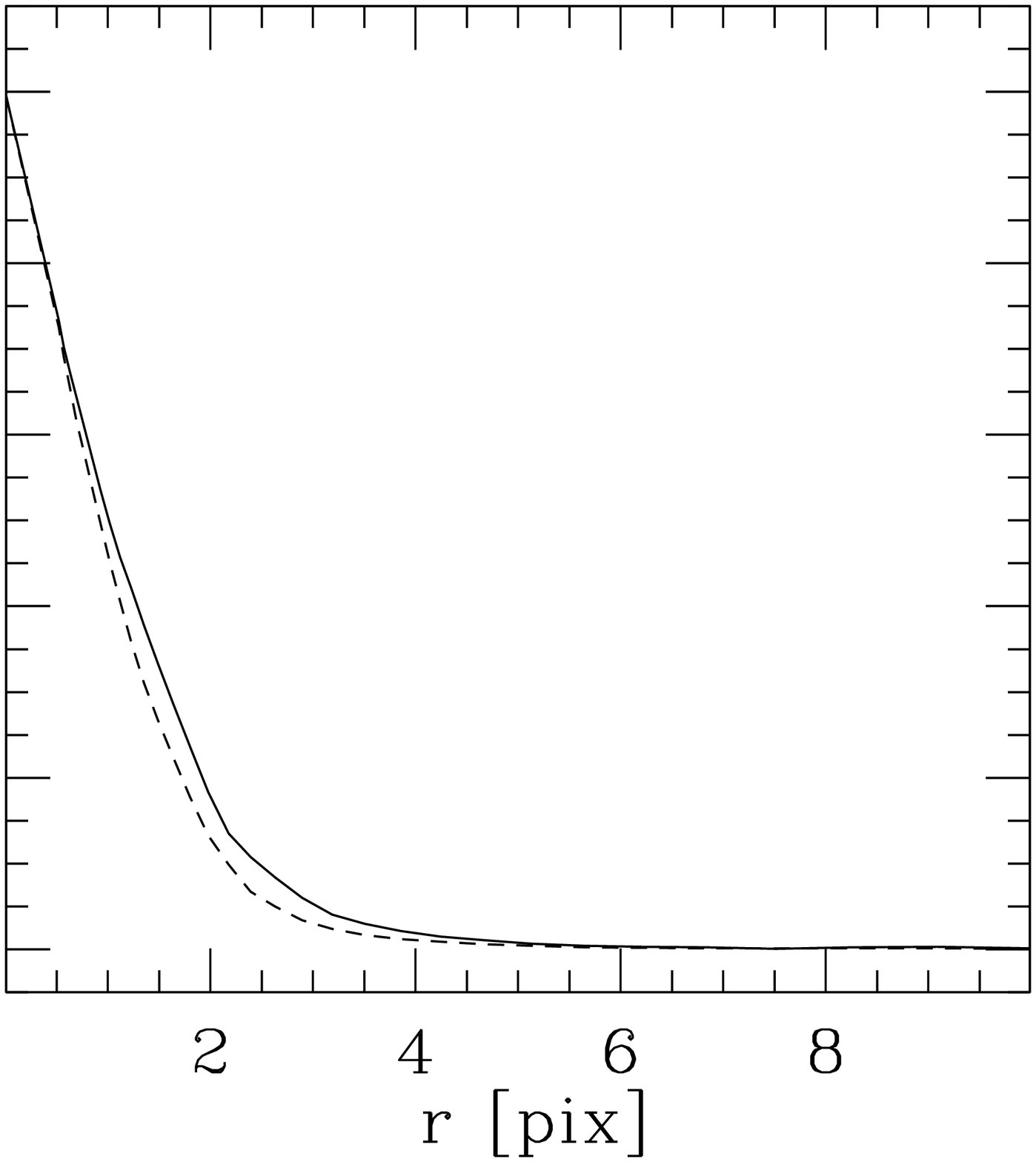,width=4.2cm}

\caption{{\bf Top:} Two $3''\times 2.5''$ HST archival image excerpts of Centaurus compact objects (a total of 9 objects have HST imaging available, see text). 
  The images are from an ACS image in the F435W filter (Proposal 9427,
  PI Harris) with matched intensity cuts. Object CCOS J1248.74-4118.58
  is the largest source of the sample, object CCOS J1248.76-4118.70
  the smallest one (see also Fig.~\ref{sizes}). {\bf Bottom:}
  Arbitrarily normalised radial intensity profile of both sources,
  compared to a stellar PSF (dashed line).}
\label{ucdthumbs}
\end{center}
\end{figure}

\begin{figure}[]
\begin{center}
  \epsfig{figure=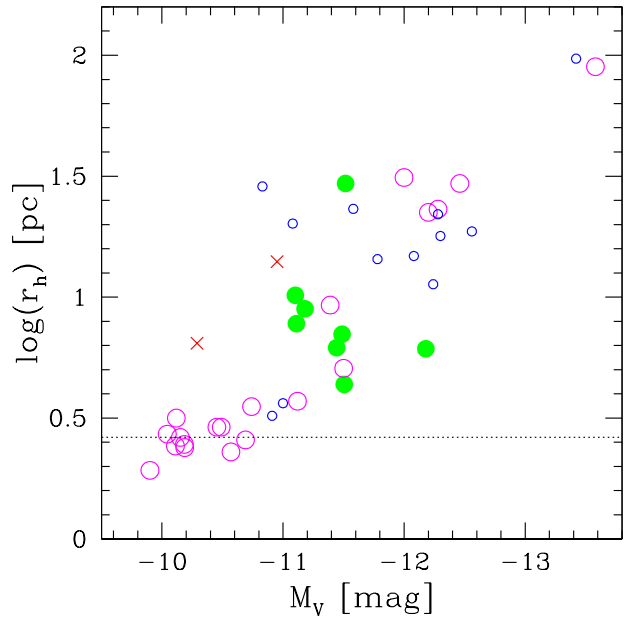,width=8.6cm}
\caption{HST based half light radius measurements of compact objects in various environments plotted vs. their absolute magnitude $M_V$. Green filled circles indicate Centaurus compact objects (this paper). The dotted line indicates the
  resolution limit of the Centaurus ACS images. This value
  ($r_h\simeq$ 2.5pc) also corresponds closely to the average GC size.
  Magenta open circles indicate Fornax compact objects from
  Mieske et al.~(\cite{Mieske06a}) and Evstigneeva et
  al.~(\cite{Evstig07}). Blue open circles are Virgo UCDs from
  Evstigneeva et al.~(\cite{Evstig07}) and Ha\c{s}egan et
  al.~(\cite{Hasega05}).  Red crosses are the Local Group compact
  objects $\omega$Cen (fainter) and G1 (brighter).}
\label{sizes}
\end{center}
\end{figure}

Out of the 27 compact objects detected in the CCOS, eight have
archival HST imaging available. Six objects are found in one ACS
pointing of NGC 4696 (Proposal 9427, PI Harris), while the two
remaining sources are contained in two WF chips of one WFPC2 pointing
(Proposal 6579, PI Tonry). Two example thumbnails are shown in
Fig.~\ref{ucdthumbs}. We use these images to estimate their sizes.  To
this end we use the program KINGPHOT (Jord\'an et al.~\cite{Jordan04}
and~\cite{Jordan05}), which was already successfully applied to
measure half-light radii $r_h$ of GCs in Virgo and Fornax (Jord\'an et
al.~\cite{Jordan05}~and~\cite{Jordan07}).  At the distance of the
Centaurus cluster (assumed to be 43 Mpc), the resolution limit of the
ACS images in terms of half light radius corresponds to $\sim$2.5 pc,
i.e. about the typical value of GC $r_h$.  The limit of the WF images is $\sim$
5 pc. The resulting $r_h$ estimates are listed in Table~\ref{table}.
Fig.~\ref{sizes} plots $M_V$ vs. $r_h$ in pc measured on the HST
images for the eight Centaurus compact objects.  Over-plotted are HST
based size estimates for compact objects in other environments:
Fornax, Virgo, and the Local Group. The largest Centaurus compact
object (CCOS J1248.74-4118.58) has a half-light radius of about
30pc.  Seven further objects have much smaller sizes in the range 4 to
10 pc, which is only marginally larger than typical GC sizes.

\section{Discussion}
\label{discussion}
\subsection{Can we identify a bona-fide UCD sub-sample?}
In several respects (luminosity distribution, total number, spatial
distribution, kinematics) the compact object sample does not show
significant differences compared to what is expected from the globular
cluster population. That is, based on those properties we are not able
to identify a separate sub-population of UCDs. What can we say
about our compact object sample regarding sizes and abundances?

\subsubsection{Sizes}

Only one of the nine compact objects with HST imaging has a large size
of $r_h=$30 pc that clearly separates it from the typical range of GC
sizes. This object CCOS J1248.74-4118.58 truly classifies as an
ultra-compact dwarf galaxy. In logarithmic space, its size is right
between those of GCs and dwarf spheroidal galaxies (Mateo et
al.~\cite{Mateo98}).

The overall size distribution of the Centaurus compact objects fits
into the picture outlined in Ha\c{s}egan et al.~(\cite{Hasega05}) and
Mieske et al.~(\cite{Mieske06a}): for $M_V<-11$ mag the sizes of
compact objects start to deviate from the typical GC values towards
larger sizes. This may indicate that those large sources are the
merger of several single star clusters (Kissler-Patig et
al.~\cite{Kissle05}, Bekki et al.~\cite{Bekki04}, Fellhauer \&
Kroupa~\cite{Fellha02}). The magnitude range of our compact objects is
too small and their size scatter too large to confirm a {\it trend} of
increasing size with luminosity, as it is found for Fornax and Virgo. This is
partially due to the fact that we do not detect a very bright and
large UCD with $M_V\simeq$ -13.5 mag as in Fornax and Virgo. However,
our survey completeness of 30\% still allows at 95\% confidence for
3-4 sources in the bright magnitude regime where no compact objects
are found. We have to substantially increase our survey completeness
to make firm statements on the size (and mass) range of UCDs in Centaurus.


\subsubsection{Abundances}
Our Centaurus sample does not show any obvious break in the
metallicity distribution (Fig.~\ref{Fe_V2}) that would allow a
sub-division in two groups of objects. The sample has a metallicity
distribution whose mean agrees very well with the value of Fornax UCDs
and that of the metal-rich GC subpopulation, with the scatter being
somewhat larger than in Fornax.  We note that our survey has an
absolute magnitude limit of $M_V=-10.9$ mag, such that we would not be
able to trace a metallicity break at $M_V=-11$ mag like that found for Fornax
compact objects (Mieske et al.~\cite{Mieske06a}).

There may be a dichotomy in $\alpha$ abundance between bright and
faint Centaurus compact objects, but this certainly needs confirmation
with higher S/N data. Even if confirmed, this would not necessarily
imply different formation mechanisms for bright and faint sources. The
recently found colour-luminosity relation among bright blue GCs in
various nearby GC systems (e.g.  Harris et al.~\cite{Harris06},
Strader et al.~\cite{Strade06}, Mieske et al.~\cite{Mieske06b})
indicate that one may have abundance trends within one GC population.
The findings of multiple or broadened main sequences in some massive
Milky Way GCs (Bedin et al.~\cite{Bedin04}; D'Antona et
al.~\cite{Danton05}; Piotto et al.~\cite{Piotto07}) also point to the
possibility of having complex star formation histories within massive GCs.

\vspace{0.4cm}

\noindent We note that another potential separation criterion between GCs
and UCDs are M/L ratios (Ha\c{s}egan et al.~\cite{Hasega05}).
However, our low-resolution spectra do not allow to measure intrinsic
velocity dispersions. This would require several nights of on-source
integration at significantly higher spectral resolution.

\subsection{Clues to the formation of UCDs}

The only bona-fide UCD detected in our survey appears to have a
super-solar [$\alpha$/Fe] abundance, different to typical values found
for the current population of dEs. When including the abundance
analyses done for UCD populations in Fornax and Virgo, it thus appears
that today's dwarf elliptical galaxy population is not the parent
population of most UCDs.  The stellar super cluster scenario (SSC)
(e.g.  Fellhauer \& Kroupa~\cite{Fellha02}) is a conceivable
possibility for the formation of UCDs, also because young and very
massive super-clusters -- the possible UCD progenitors -- can be
observed in various environments today (e.g.  Kissler-Patig et
al.~\cite{Kissle05}, Bastian et al.~\cite{Bastia05}, Cortese et
al.~\cite{Cortes07}). One would not expect too metal-poor SSCs given
that they are created by already pre-processed material. This fits to
the metallicities estimated for most of the UCDs. The scenario seems
especially plausible for the Centaurus cluster, which is believed to
have had a quite busy merger history (e.g. Furusho et
al.~\cite{Furush01}).

\section{Conclusions}
In this paper we have presented a search for ultra-compact dwarf
galaxies in the Centaurus galaxy cluster. We acquired spectroscopic
redshifts of about 400 compact object candidates with $19.2<V<22.4$
mag ($-14.1<M_V<-10.9$ mag at the distance of Centaurus) with
VIMOS@VLT.  We find 27 compact objects (21 in Cen30, 6 in Cen45) with
radial velocities consistent with Centaurus, covering an absolute
magnitude range $-12.2<M_V<-10.9$ mag.

Most properties (e.g. number, luminosity distribution, spatial
distribution) of the compact object sample can be explained by the
globular cluster systems in Centaurus, such that it is difficult to
identify a UCD sub-sample.  Only one of the eight sources with HST
imaging has a large size of $r_h\sim$ 30 pc which clearly
distinguishes it from normal globular clusters. We consider this
source the only bona-fide UCD detected in our survey.  It has an
$\alpha$ abundance well above that of typical dwarf elliptical
galaxies, more in line with Galactic bulge globular clusters. For this
source, creation as a merged stellar super cluster in a past galaxy
merger event appears plausible. To further quantify the properties of
Centaurus UCDs, a significant increase of our survey completeness is
necessary.

\label{conclusions}

\begin{table*}
\caption{Properties of the 27 Centaurus compact objects detected in our survey, ordered by magnitude. Errors are given in parentheses. ``CCOS'' in the object identifier stands for Centaurus Compact Object Survey. The last column gives the half-light radius in pc estimated from HST archival imaging. }
\label{table}
\begin{tabular}{l|rrrrlrrr}
ID & RA (J2000) & DEC (J2000)&V$_0$ & (V-R)$_0$& v$_{\rm rad}$ [km/s] & $\langle {\rm Fe}\rangle $ [\AA] & Mgb [\AA] & r$_h$ [pc]\\\hline\hline
CCOS J1250.02-4121.45 & 12:50:01.23 & -41:21:26.83 & 21.06 & 0.40 & 4835 (60) & 2.76 (0.36) & 4.13 (0.55) & \\ 
CCOS J1248.54-4119.64 & 12:48:32.64 & -41:19:38.20 & 21.08 & 0.47 & 3854 (69) & 2.20 (0.27) & 2.83 (0.36) & \\ 
CCOS J1248.79-4117.15 & 12:48:47.61 & -41:17:09.12 & 21.10 & 0.50 & 2882 (54) & 4.15 (0.25) & 6.69 (0.36) & 6.12 (0.45)\\ 
CCOS J1248.06-4111.40 & 12:48:03.76 & -41:11:24.30 & 21.15 & 0.58 & 2768 (89) & 2.80 (0.20) & 2.96 (0.27) & \\ 
CCOS J1248.54-4121.90 & 12:48:32.48 & -41:21:54.07 & 21.22 & 0.51 & 3279 (75) & 3.51 (0.30) & 4.07 (0.42) & \\ 
CCOS J1248.70-4121.11 & 12:48:41.75 & -41:21:06.76 & 21.57 & 0.50 & 2521 (123) & 3.56 (0.40) & 2.78 (0.57) & \\ 
CCOS J1250.11-4114.59 & 12:50:06.57 & -41:14:35.28 & 21.57 & 0.45 & 4430 (66) & 2.70 (0.35) & 2.44 (0.47) & \\ 
CCOS J1249.71-4120.63 & 12:49:42.63 & -41:20:38.04 & 21.59 & 0.43 & 5184 (55) & 2.29 (0.35) & 3.34 (0.48) & \\ 
CCOS J1248.79-4116.00 & 12:48:47.69 & -41:16:00.11 & 21.75 & 0.46 & 3140 (63) & 3.73 (0.28) & 5.54 (0.40) & \\ 
CCOS J1248.74-4118.58 & 12:48:44.70 & -41:18:35.03 & 21.76 & 0.44 & 2189 (114) & 1.34 (0.59) & 3.35 (0.74) & 29.49 (0.64)\\ 
CCOS J1248.76-4118.70 & 12:48:45.56 & -41:18:42.21 & 21.77 & 0.55 & 2828 (45) & 4.15 (0.47) & 6.09 (0.68) & 4.36 (0.30)\\ 
CCOS J1248.74-4119.74 & 12:48:44.33 & -41:19:44.37 & 21.79 & 0.40 & 3138 (56) & 4.28 (0.35) & 5.77 (0.49) & 7.02 (0.25)\\ 
CCOS J1250.03-4122.68 & 12:50:01.77 & -41:22:40.60 & 21.84 & 0.54 & 4538 (66) & 0.96 (0.61) & 3.53 (0.85) & 6.18 (0.83)\\ 
CCOS J1248.63-4115.71 & 12:48:37.77 & -41:15:42.37 & 21.87 & 0.39 & 2524 (88) & 2.61 (0.47) & 2.40 (0.65) & \\ 
CCOS J1249.07-4120.77 & 12:49:03.99 & -41:20:45.96 & 21.88 & 0.37 & 2453 (140) & 0.87 (0.39) & 3.24 (0.51) & \\ 
CCOS J1249.36-4121.57 & 12:49:21.64 & -41:21:34.16 & 22.09 & 0.42 & 2790 (96) & 2.44 (0.58) & 4.76 (0.76) & \\ 
CCOS J1249.99-4124.45 & 12:49:59.21 & -41:24:27.18 & 22.10 & 0.36 & 5000 (80) & 1.47 (0.62) & 3.59 (0.88) & 8.94 (1.31) \\ 
CCOS J1248.97-4120.01 & 12:48:58.25 & -41:20:00.52 & 22.12 & 0.52 & 2962 (98) & 1.88 (0.45) & 5.89 (0.62) & \\ 
CCOS J1248.65-4120.34 & 12:48:39.28 & -41:20:20.58 & 22.13 & 0.55 & 3303 (69) & 3.55 (0.53) & 4.88 (0.76) & \\ 
CCOS J1248.74-4118.95 & 12:48:44.11 & -41:18:56.87 & 22.17 & 0.53 & 3213 (53) & 2.35 (0.65) & 4.66 (0.90) & 7.77 (1.32)\\ 
CCOS J1248.70-4118.23 & 12:48:41.99 & -41:18:13.77 & 22.18 & 0.48 & 2822 (55) & 3.47 (0.43) & 5.96 (0.60) & 10.17 (0.98)\\ 
CCOS J1250.07-4121.16 & 12:50:04.49 & -41:21:09.37 & 22.18 & 0.43 & 4952 (80) & 0.09 (0.92) & 0.72 (1.25) & \\ 
CCOS J1248.72-4121.23 & 12:48:43.02 & -41:21:13.54 & 22.23 & 0.58 & 2582 (99) & 2.26 (0.60) & 3.21 (0.86) & \\ 
CCOS J1249.41-4118.64 & 12:49:24.85 & -41:18:38.15 & 22.23 & 0.58 & 2206 (79) & 3.42 (0.60) & 5.58 (0.76) & \\ 
CCOS J1248.91-4117.70 & 12:48:54.64 & -41:17:42.05 & 22.29 & 0.43 & 2909 (120) & 1.78 (0.67) & 3.93 (0.88) & \\ 
CCOS J1248.99-4118.08 & 12:48:59.65 & -41:18:04.51 & 22.31 & 0.53 & 2806 (75) & 2.15 (0.66) & 4.72 (0.91) & \\ 
CCOS J1248.62-4120.63 & 12:48:37.13 & -41:20:37.79 & 22.34 & 0.52 & 3741 (85) & 2.96 (0.61) & 4.48 (0.87) & \\ 
\hline
\end{tabular}
\end{table*}

\begin{table*}
\caption{Properties of the 13 Centaurus cluster dEs observed in our study. The first column gives the identifier from the Centaurus cluster catalog (CCC, Jerjen et al.~\cite{Jerjen97}). The B$_{\rm T}$ magnitude is from the CCC, all other observables are from this study. The last column gives the radial velocity measured for 9 of the 13 sources by Stein et al.~(\cite{Stein97}).}
\label{table_dE}
\begin{tabular}{l|rrrrlrrr}
ID & RA (J2000) & DEC (J2000)&B$_{\rm T}$ & v$_{\rm rad}$ [km/s] & $\langle {\rm Fe}\rangle $ [\AA] & Mgb [\AA] & v$_{\rm rad, S97}$ [km/s] \\\hline\hline
CCC  125 & 12:49:56.31 & -41:15:35.97 & 17.14 & 3016 (78) & 2.85 (0.34) & 4.18 (0.45) & 2880 (35)\\ 
CCC   61 & 12:48:39.68 & -41:16:03.14 & 17.40 & 2871 (62) & 3.72 (0.12) & 4.14 (0.17) & 2910 (67) \\ 
CCC   58 & 12:48:36.20 & -41:26:21.38 & 17.94 & 3312 (58) & 3.20 (0.23) & 3.65 (0.33) &  3304 (60) \\ 
CCC   22 & 12:48:02.05 & -41:18:18.25 & 17.94 & 2467 (38) & 2.79 (0.09) & 3.90 (0.13) & 2433 (33)\\ 
CCC   75 & 12:49:02.05 & -41:15:33.26 & 18.06 & 1977 (73) & 3.34 (0.19) & 3.62 (0.26) & 1958 (71)\\ 
CCC  150 & 12:50:24.36 & -41:17:46.91 & 18.23 & 4408 (99) & 2.49 (0.15) & 3.07 (0.20) &  4426 (46) \\ 
CCC  104 & 12:49:35.82 & -41:25:36.59 & 18.25 & 3317 (75) & 3.54 (0.47) & 4.89 (0.65) & --- \\ 
CCC  121 & 12:49:54.15 & -41:20:21.39 & 18.36 & 4661 (62) & 1.55 (0.30) & 2.96 (0.41) & 4739 (70)  \\ 
CCC   52 & 12:48:29.99 & -41:19:16.93 & 18.42 & 3768 (73) & 3.59 (0.27) & 4.05 (0.37) & --- \\ 
CCC  123 & 12:49:56.03 & -41:24:04.06 & 18.45 & 4729 (46) & 2.42 (0.14) & 2.75 (0.20) & 4661 (69) \\ 
CCC   97 & 12:49:30.19 & -41:25:14.93 & 18.70 & 2870 (70) & 3.58 (0.19) & 4.63 (0.25) & 2818 (30) \\ 
CCC   72 & 12:48:58.81 & -41:10:26.55 & 19.11 & 3218 (54) & 2.74 (0.27) & 2.20 (0.37) & ---\\ 
CCC   16 & 12:47:54.58 & -41:18:43.37 & 19.40 & 3891 (58) & 2.78 (0.19) & 3.27 (0.27) & --- \\ \hline
\end{tabular}
\end{table*}
\end{document}